# Time-Temperature-Transformation (TTT) Diagrams to rationalize the nucleation and quenchability of metastable α-Li$_3$PS$_4$

Akira Miura,[a†]* Woohyeon Baek,[b†] Yuta Fujii,[a] Kiyoharu Tadanaga,[a] Rana Hossain,[a] Yamashita Aichi,[c] Yoshikazu Mizuguchi,[c] Chikako Moriyoshi,[d] Shintaro Kobayashi,[e] Shogo Kawaguchi,[e] Ding Jiong,[f] Shigeo Mori,[f] Atsushi Sakuda,[g] Akitoshi Hayashi,[g,h] Wenhao Sun[b]*

[a] Faculty of Engineering, Hokkaido University, Kita 13, Nishi 8, Sapporo, 060-8628, Japan

[b] Department of Materials Science and Engineering, University of Michigan, Ann Arbor, MI, 48109, USA

[c] Department of Physics, Tokyo Metropolitan University, Hachioji, 192-0397, Japan

[d] Graduate School of Advanced Science and Engineering, Hiroshima University, 1-3-1, Kagamiyama, Higashihiroshima 739-8526, Japan

[e] Japan Synchrotron Radiation Research Institute, Sayo, Hyogo, 679-5198, Japan

[f] Department of Materials Science, Graduate School of Engineering, Osaka Metropolitan University, 1-1 Gakuen-cho, Naka-ku, Sakai, Osaka, 599-8531, Japan

[g] Department of Applied Chemistry, Graduate School of Engineering, Osaka Metropolitan University, 1-1 Gakuen-cho, Naka-ku, Sakai, Osaka, 599-8531, Japan

[h] Institute for Materials Research, Tohoku University, 2-1-1 Katahira, Aoba-ku, Sendai, Miyagi 980-8577, Japan

† contributed equally    Email: amiura@eng.hokudai.ac.jp (AM), whsun@umich.edu (WS)

**Abstract**: α-Li$_3$PS$_4$ is a promising solid-state electrolyte with the highest ionic conductivity among its polymorphs. However, its formation presents a thermodynamic paradox: the α-phase is the equilibrium phase at high temperature and transforms to the stable γ-Li$_3$PS$_4$ polymorph when cooled to room temperature; however, α-Li$_3$PS$_4$ can be synthesized and quenched in a metastable state via rapid heating at relatively low temperatures. The origin of this synthesizability and anomalous stability has remained elusive. Here, we resolve this paradox by establishing a comprehensive time-temperature-transformation (TTT) diagram, constructed from a computational temperature-size phase diagram and experimental high-time-resolution isothermal measurements. Our density functional theory calculations reveal that at the nanoscale, the α-phase is stabilized by its low surface energy, which drastically lowers the nucleation barrier across a wide temperature range. This size-dependent stabilization is directly visualized using in-situ synchrotron X-ray diffraction and electron microscopy, capturing the rapid nucleation of nano-sized α-phase and its subsequent slow transformation. This work presents a generalizable framework that integrates thermodynamic and kinetic factors for understanding nucleation and phase transformation mechanisms, providing a rational strategy for the targeted synthesis of functional metastable materials.

Metastable materials, ranging from traditional steel and glass to advanced battery materials, are crucial for a wide range of functional applications.[1-3] Numerous synthesis techniques have been developed to obtain metastable materials, including topotactic intercalation,[1] ball milling,[2] and rapid temperature control.[3] Among these, controlling the heating profile remains a key strategy for producing metastable phases with desired properties. For instance, the classic composition-temperature phase diagram of the Fe-C system shows thermodynamically stable phases, while the temperature-time-transformation (TTT) diagram provides kinetic guidance for processing both stable and metastable phases in steel.[4]

Multidimensional phase diagrams, which extend beyond conventional two-dimensional temperature-pressure or temperature-composition diagrams, can be powerful tools for capturing the thermodynamics and processing conditions for producing metastable materials.[5] By combining first-principles theories and computations, we can achieve a predictive understanding to guide the synthesis and processing pathways of specific metastable phases.[6,7] For example, DFT-computed size-dependent phase diagrams can explain the nanoscale stability of metastable manganese oxides,[8,9] or competitive nucleation of Sc-Zn quasicrystals by accounting for time- and temperature-dependent phase selectivity.[10] Chemical potential diagrams of multicomponent materials have also been proposed to understand the chemical reaction pathway to target materials.[11,12]

The theory of TTT diagrams is well-established for metallurgical and glass systems.[4,13] Recently, the TTT diagram have been expanded to the electronic system.[14] However, in functional ceramics, such as battery materials, TTT diagrams are typically obtained experimentally,[15,16] and there is currently not a strong theoretical foundation to predict or interpret TTT diagrams for ceramics. Moreover, experimentally measured TTT diagrams are usually constructed from thermal analysis (e.g., TG-DTA) or ex-situ XRD characterization of samples subjected to different heating protocols, which might not be able to resolve distinct thermal events if they convolve into one signal during fast nucleation and transformation processes. Furthermore, ex-situ XRD lacks the necessary time resolution to capture the rapid reactions that occur within seconds, making it difficult to understand the early stages of structure selection.[12]

Recent advances in in-situ synchrotron X-ray techniques have enabled data acquisition with high time resolution, offering a powerful perspective for understanding metastable phase formation during reactions. By combining a preheated atmosphere with a rapidly inserted sample holder, along with rapid heating and subsequent isothermal heating experiments, we and others[17] have now developed the capability to capture the fast, short timescale formation of metastable phases. These experiments provide essential kinetic insights into their metastable nature, and the origin of phase selectivity depending on synthesis protocol.[17,18]

One recent mystery in the temperature-time processing of metastable phases is the quenchability of α-$Li_3PS_4$. Lithium thiophosphate ($Li_3PS_4$) is a well-established solid sulfide electrolyte for all solid-state batteries, and has four polymorphs: amorphous, low-temperature γ phase, medium-temperature β phase, and high-temperature α phase.[19,20] The α phase has the highest Li-ionic conductivity for solid-state electrolytes, but is only thermodynamically stable at temperatures above 480°C, and typically converts to the less ionically-conductive γ phase upon cooling.[19] Recently, it was observed that the α phase can be synthesized by rapid thermal annealing (400 °C/min) of the amorphous glass precursor at 243-374 °C followed by quenching; whereas a longer-term heating produces the *β* phase.[3] Why can the high-temperature polymorph α-$Li_3PS_4$ form at low temperatures, and more importantly, why can it be retained when quenched quickly? If we could build a thermodynamic and kinetic understanding of this process, it could broadly enable the construction of TTT diagrams to synthesize and retain high-performance metastable materials.

In this study, we provide a comprehensive theoretical and experimental understanding of the stability and transformation kinetics in the $Li_3PS_4$ system. Computational analysis of bulk and surface energies show that α-$Li_3PS_4$ is bulk metastable, but is stabilized at the nanoscale over a broad temperature range by its low surface energy. This offers a starting point to understand why its nanocrystalline form is quenchable to room temperature, in contrast to its bulk counterpart, which transforms over time to γ-$Li_3PS_4$. Next, using in-situ synchrotron XRD with millisecond time resolution, we reveal the kinetic persistence of this low-temperature metastable pathway, where rapid nucleation plus inhibited crystal growth allows nanoscale α-$Li_3PS_4$ to minimize its driving force for transformation to the equilibrium γ-phase. Our combined approach provides a holistic view of the thermodynamic and kinetic factors governing phase selection in metastable functional ceramics.

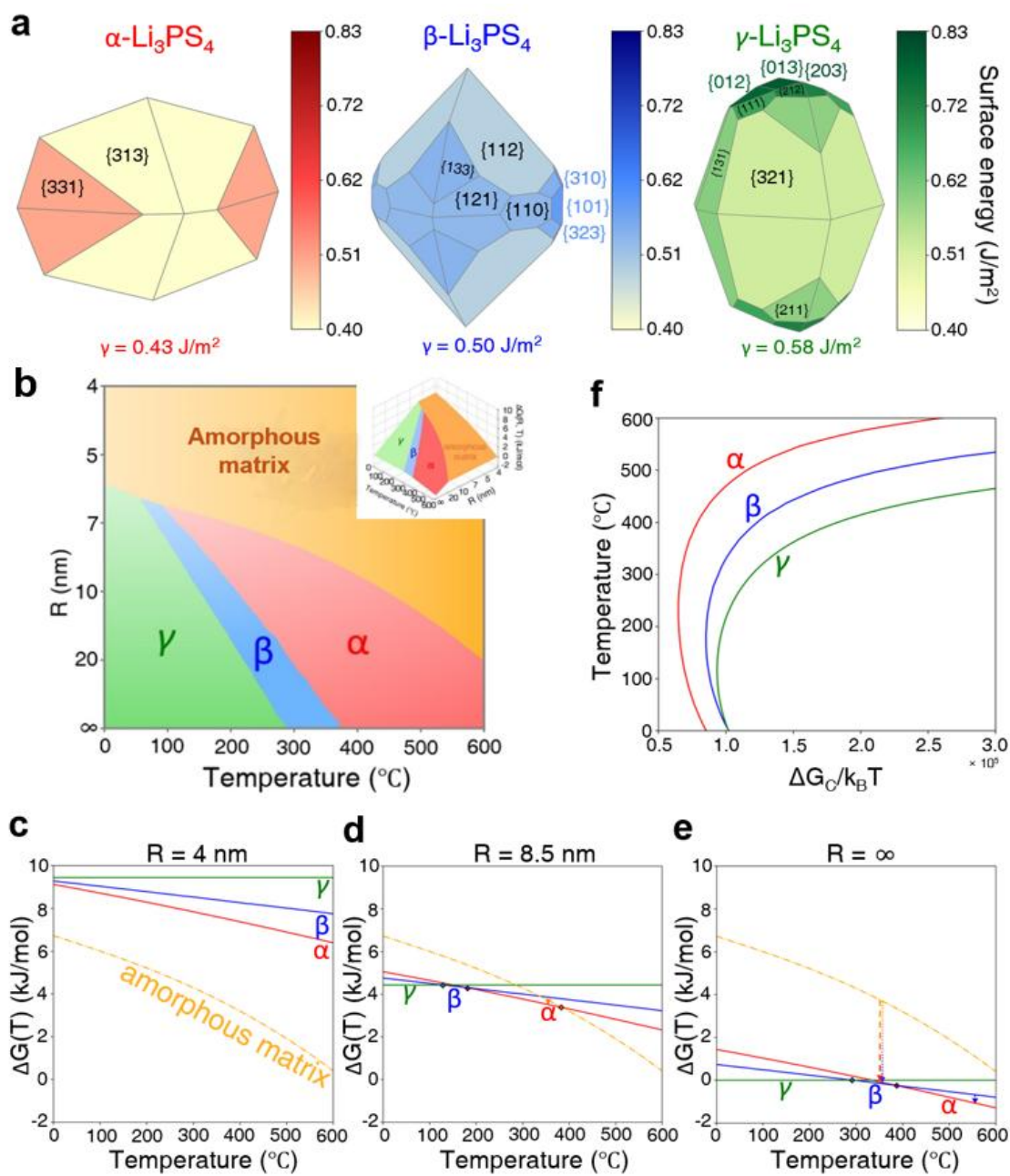

**Figure 1. Temperature- and size-dependent phase stability of Li$_3$PS$_4$ polymorphs from first-principles calculations. a,** Calculated Wulff shapes of α-, β-, and γ-Li$_3$PS$_4$, with facets colored according to their surface energy. **b,** The three-dimensional and projected two-dimensional phase diagram showing the Gibbs free energy, ΔG(T, R), as a function of temperature (T) and particle radius (R). The colored surfaces represent the most thermodynamically stable phase in each region. **c-e,** Two-dimensional plots of Gibbs free energy versus temperature, showing cross-sections of the 3D diagram for particle radii of **c,** R = 4 nm, **d,** R = 8.5 nm, and **e,** the bulk limit (R = ∞). All energies are referenced to bulk γ-Li$_3$PS$_4$ (R = ∞) at each temperature. Diamond points are the transition point where the free energy of phases is the same. Arrows

are the driving force to transform from amorphous to nanosized α-Li$_3$PS$_4$ at 350 °C (dashed arrow), and to β-Li$_3$PS$_4$ at 350 °C and from β-Li$_3$PS$_4$ to bulk α-Li$_3$PS$_4$ at 547 °C (dotted arrow), respectively (**Figure 2, 3, and 4**). **f,** Calculated nucleation energy barrier at the critical nucleus (Δ$G_c$) of bulk Li$_3$PS$_4$ polymorphs from the bulk amorphous matrix at different temperatures.

First, we performed *ab initio* calculations to map the phase stability of Li$_3$PS$_4$ polymorphs as a function of both temperature and particle size. We computed surface energies using our surface slab generation scheme[21] and performed density functional theory (DFT) calculations using the r$^2$SCAN-rVV10 meta-GGA functional.[22,23] From the equilibrium Wulff shapes, we calculated the morphology-averaged surface energies of the crystalline polymorphs, shown in **Figure 1a** (Methods).[24,25] The morphology-averaged surface energy is lowest for α-Li$_3$PS$_4$ (0.43 J/m$^2$), intermediate for the β phase (0.50 J/m$^2$), and highest for the γ phase (0.58 J/m$^2$). This ordering is the inverse of the calculated bulk formation free energy at ambient temperatures ($G_\gamma < G_\beta < G_\alpha$), meaning that there is an opportunity for a nanoscale crossover in phase stability,[26,27] where the bulk metastable α-phase can be nano-stabilized at high surface area to volume ratios (**Table S4**).

The combination of temperature and size variations is captured in the free energy surfaces presented in **Figure 1b**. The vibrational and configurational entropy contributions of the α-, β- and γ- polymorphs were calculated using quasi-harmonic phonons[28,29,30,31] with Wang-Landau sampling of the vibrational density of states[32] (Methods, Supplementary Information, and Figure S2). For the bulk polymorph ordering (R = ∞), the calculated phase transition temperatures are 296 °C for γ→β; and 385 °C for β→α (**Figure 1e**), which are in good agreement with experimental measurements (250–300 °C and 450–500 °C, respectively).[3,19,20] This behavior changes dramatically for nanoparticles, as shown in the three-dimensional phase diagram (**Figure 1b**) and its two-dimensional cross-sections (**Figure 1c, d, e**). Here, the nanoscale free energy of crystalline phases are calculated from size-dependent energy relations $G(T,R) = H - TS_{vib} + \gamma\eta\rho(1/R)$, where $\gamma$ the surface area, $\rho$ is the atomic density and $\eta$ is the shape factor.[10] At a particle radius of 8.5 nm, the γ→β and β→α transition temperatures are suppressed to 127 °C and 180 °C, respectively (**Figure 1d**).

The crystalline polymorphs nucleate from an amorphous matrix, and so we also need the free energy of the amorphous phase. Unfortunately, the enthalpy and entropy of amorphous phases are difficult to obtain from DFT, as phonons are difficult to converge. Here we introduce a scheme that references the free energy of the amorphous phase from experimental differential scanning calorimetry (DSC) data of the amorphous → β transition at 230°C (**Figure S2d**).[3] The DSC-measured enthalpy evolution for Δ$H_{amorph→β}$ = -4.8 kJ/mol

(Methods and Supplementary Information), and the $\Delta S$ between $\beta \rightarrow$ amorphous can be determined by integrating the measured heat capacity for each phase in the temperature window near the phase transition. The calculated free energy of the bulk amorphous phase, derived from experimental data, was then included into the phase diagram (**Figure 1b, c, d, e**). The free energy of amorphous matrix is fixed to be the bulk energy (*i.e.* we do not evaluate its size-independent free energy) because the amorphous phase is the precursor phase, and is only in a bulk state in this scenario. At the nanoscale, the bulk amorphous phase crosses over in free energy first to crystalline α-Li₃PS₄ at a radius of 8.5 nm and 380 °C (**Figure 1d**) and at bulk sizes is metastable with respect to all crystalline phases (**Figure 1e**).

Finally, to address the nucleation kinetics of polymorph selectivity, the critical nucleation barrier, $\Delta G_c$, was calculated for each crystalline polymorph from a parent bulk amorphous matrix (**Figure 1f and S3**). The steady-state nucleation rate is obtained from $J \sim \exp(-\Delta G_c/k_B T)$, where $\Delta G_c = 4/27 \ [(\eta \gamma)^3 / (\Delta G_{bulk})^2]$, where $\Delta G_{bulk}$ is the free energy difference between the parent amorphous matrix and the various Li₃PS₄ polymorphs. We plot $\Delta G_c/k_B T$ as a function of temperature in **Figure 1f**. Because the α-phase is the polymorph with the lowest surface energy, it also has the lowest nucleation barrier at all temperatures. A characteristic 'nose' shape appears in this $\Delta G_c/k_B T$ diagram, where α has the smallest barrier at 227 °C, due to a trade-off between increasing $\Delta G_{bulk}$ driving force at low temperatures, but being divided by $k_B T$ at low temperatures. Overall, **Figure 1f** indicates that the α-phase can nucleate first at a significantly lower temperature than the β- and γ-phases.

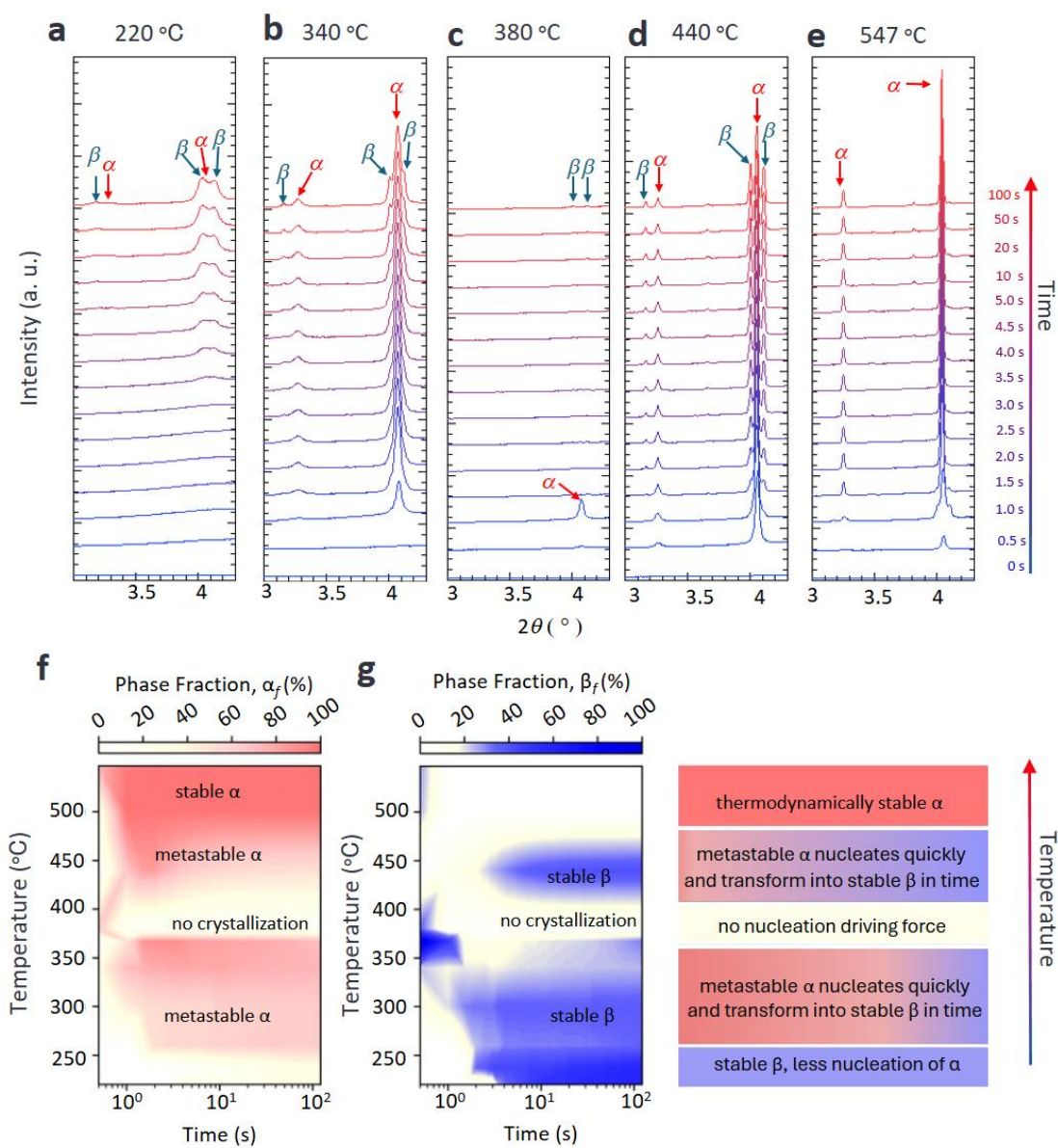

**Figure 2. Experimental mapping of the time- and temperature-dependent phase evolution of Li$_3$PS$_4$ from an amorphous precursor.** Time-resolved XRD patterns with the wavelength of 0.354 nm collected during isothermal annealing at **a,** 220 °C, **b,** 340 °C, **c,** 380 °C, **d,** 440 °C, and **e,** 547 °C. The amorphous sample was rapidly heated to the target temperature (before t=0), and the subsequent phase evolution was monitored over time. Red and blue arrows indicate characteristic peaks for the α- and β-phases, respectively. **f, g,** Time-Temperature-Transformation (TTT) diagrams constructed from the in-situ XRD data. The diagram is divided into two halves: the left side (red colormap) displays the phase fraction of α-Li$_3$PS$_4$ (α$_f$), while the right side (blue colormap) displays the phase fraction of β-Li$_3$PS$_4$ (β$_f$) as a function of time and temperature.

To experimentally map the crystallization kinetics, we employed high-resolution in-situ synchrotron XRD at the SRring-8 beamline, where we are able to achieve 0.5 sec-time resolution. The time-resolved diffraction patterns (**Figure 2a-e**) and the resulting Time-Temperature-Transformation (TTT) diagram (**Figure 2**, bottom) reveal temperature-dependent transformation pathways when annealing from the amorphous precursor.

The observed crystallization behavior can be categorized into four distinct regimes. First, at low temperatures around 220 °C (**Figure 2a**), the thermodynamically favored β-phase is the dominant product, though its crystal growth at these low temperatures is slow, resulting in low crystallinity. The α-phase also appears as a very minor co-existing phase.

In the intermediate temperature range of 340 °C to 440 °C (**Figure 2b, d**), a classic kinetic-to-thermodynamic transition is observed. The α-phase consistently appears first, confirming that its nucleation is kinetically favored, and persists as a metastable intermediate. With prolonged isothermal holding, the α-phase then transforms into the β-phase, which is more thermodynamically stable for larger particle sizes. Because the rate of particle growth is highly temperature-dependent, the α→β transformation proceeds significantly more rapidly and results in higher final crystallinity at 440 °C than at 340 °C.

Third, a peculiar kinetic inhibition is observed at 380 °C (**Figure 2c**). While trace amounts of the α-phase appear fleetingly within the first second, crystallization is largely suppressed for over 100 seconds, after which only weak β-phase peaks emerge. This indicates that 380 °C represents a temperature at which the nucleation driving force for both crystalline phases from the amorphous matrix is minimal, resulting in the highest stability for the amorphous phase in this time-temperature window.

Lastly, at high temperatures, such as 547 °C (**Figure 2e**), the behavior aligns with the bulk thermodynamic predictions from our calculations. The α-phase forms rapidly and remains the sole, stable crystalline phase throughout the measurement, achieving high crystallinity. This result experimentally confirms the thermodynamic stability of α-Li$_3$PS$_4$ at elevated temperatures, as predicted by the phase diagram in Figure 1.

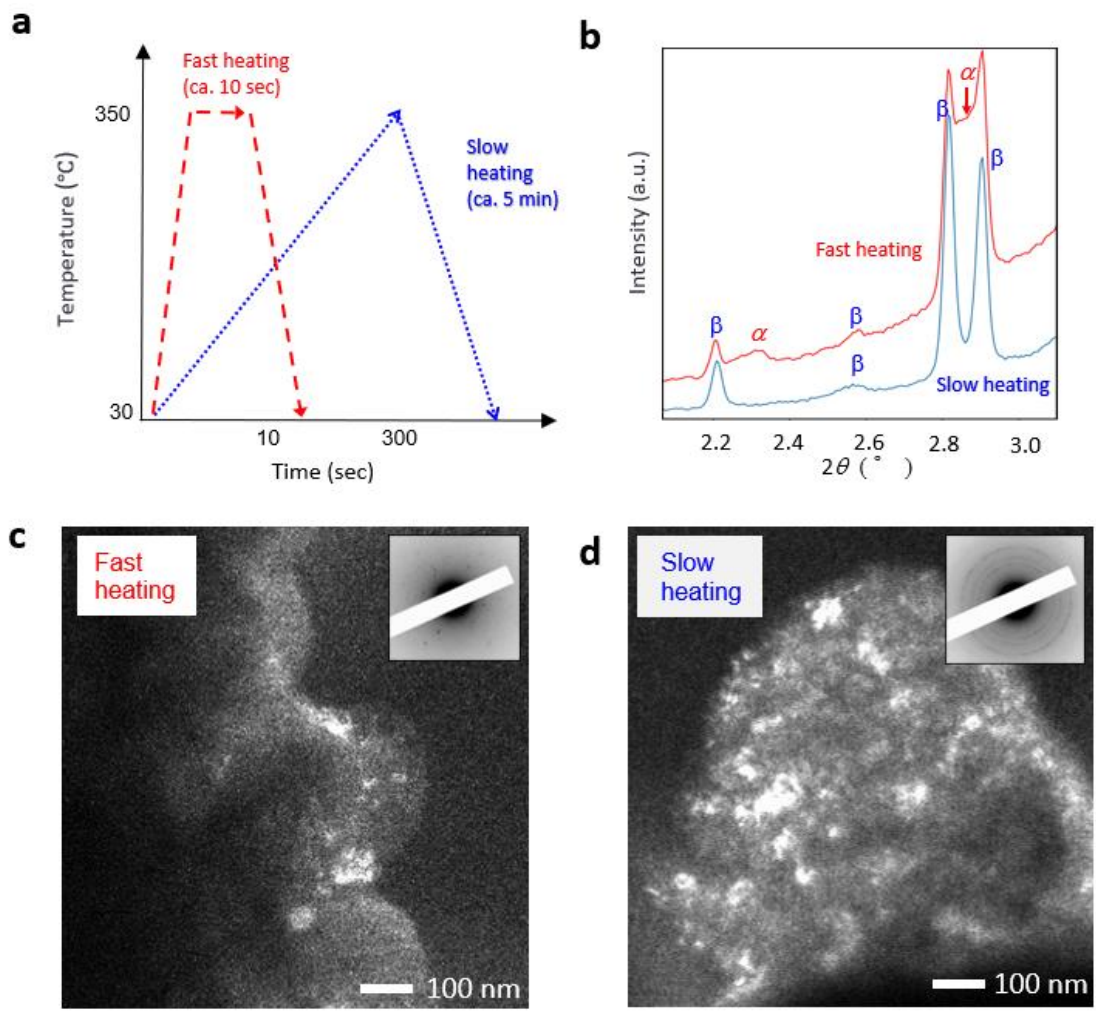

**Figure 3. Effect of heating rate on Li₃PS₄ polymorphs and microstructure after cooling to room temperature** a, Schematic illustration of two thermal profiles applied to the amorphous precursor: a fast heating process with a short hold (~10 s) and a slow heating process with an extended hold (~5 min), both to a target temperature of 350 ℃. **b,** Ex-situ XRD patterns ($\lambda = 0.248$ nm) of the final products at room temperature. The fast-heated sample exhibits a mixture of quenchable α-Li₃PS₄ and β-Li₃PS₄ phases. In contrast, the slow-heated sample shows a distinct doublet peak, a clear signature of the well-crystallized β-Li₃PS₄ phase. **c,** Dark-field TEM images, where the bright areas correspond to crystalline regions, and their corresponding selected area electron diffraction (SAED) patterns. Fast heating results in dispersed, small nanoparticles (~5-20 nm), confirmed as nanocrystalline by the diffraction spots in the SAED pattern. **d,** Slow heating leads to the formation of larger, agglomerated crystallites, consistent with the diffraction rings in SAED pattern indicating higher crystallinity. Temperature-size-energy diagram of Li₃PS₄ derived from bulk and surface energies from first-principles calculations is in Figure 1.

The effect of heating rate and quenchability of the $\alpha$-phase was examined by SXRD and TEM at room temperature. As illustrated in Figure 3a, glassy samples were heated to 350 °C using either a fast or a slow heating profile. The resulting products, analyzed by ex-situ XRD at room temperature, show distinct differences (**Figure 3b and Figure S4**). The fast-heated sample (~10 s) exhibits a mixture of $\alpha$- and $\beta$-phases, demonstrating the quenchable nature of the $\alpha$-phase. In contrast, the sample subjected to slow heating (~5 min) shows a doublet peak, which is a clear fingerprint of the well-crystallized $\beta$-phase, the equilibrium phase at this temperature. This phase formation can be explained by the TTT diagram (Figure 2), which indicates that at 350 °C with short holding times, the system is in a regime where the kinetically favored $\alpha$-phase has formed, but the transformation to the $\beta$-phase has already begun. The formed $\alpha$-phase was partially retained even after cooling to room temperature.

This phase selection is directly linked to the final microstructure, as revealed by transmission electron microscopy (TEM) analysis (**Figure 3c**). The fast-heated sample (~10 sec) consists of small, dispersed crystalline nanoparticles (5-20 nm), confirming the presence of a primary nanocrystalline phase. The slow heating profile (~5 min), however, allowed for significant crystal growth into larger agglomerated crystallites (**Figure 3d**), providing the necessary particle size for the complete transformation to the β-phase.

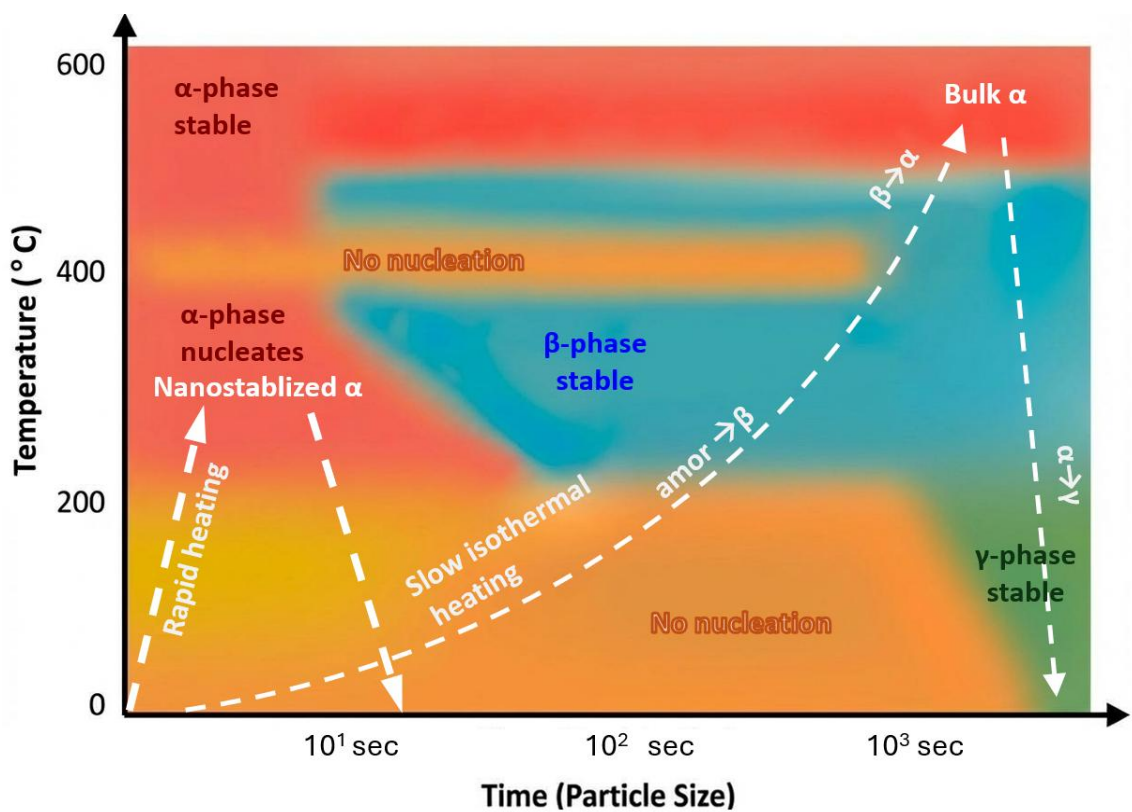

**Figure 4. A conceptual Time-Temperature-Transformation (TTT) diagram summarizing the synthesis pathways for Li$_3$PS$_4$ polymorphs.** The colored background represents the thermodynamically stable phase at a given temperature for different particle sizes (time is a proxy for particle size). The white characters and arrows illustrate two thermal pathways: (1) Rapid heating and quenching of the precursor (dashed arrow) leads to the formation of nanocrystalline α-Li$_3$PS$_4$, which can be retained at room temperature ("Quenchable α") due to nanoscale stabilization. (2) Slow isothermal heating (dotted arrow) allows for sufficient time for crystal growth, leading the system to form the more stable β-phase and eventually the high-temperature α-phase ("Unquenchable α"). Upon cooling, this bulk α-phase transforms into the stable γ-phase. This diagram provides a unified framework for understanding why the thermal history is critical for selectively synthesizing the quenchable, metastable α-phase.

The culmination of our combined computational and experimental investigation is summarized in the TTT diagram in Figure 4, which provides a unified picture of the competing thermodynamic and kinetic processes surrounding the long-standing synthesis puzzle of α-Li$_3$PS$_4$. Below 220°C and above 500°C, the synthesized products are consistent with the thermodynamically stable phases. However, in the intermediate temperatures, the small free energy differences between phases can lead to nucleation kinetics dominating structure selection. Figure 4 captures the key fact that synthesis of Li$_3$PS$_4$ is not governed by a simple temperature scale, but rather by a multidimensional landscape where thermodynamics (dependent on temperature and size) and kinetics (dependent on temperature and time) are intimately coupled.

The "Quick heating" pathway leverages the low nucleation barrier of the α-phase (**Figure 1f**) to kinetically select it over other polymorphs at low temperatures. Simultaneously, the short heating protocol prevents significant particle growth, confining the system to the nanoscale regime where the low surface energy of the α-phase (**Figure 1a**) minimizes its thermodynamic propensity to transform to the γ-phase, which is the equilibrium phase at bulk particle sizes (**Figure 1b**). This "nanoscale stabilization" is the definitive reason why the low-temperature α-phase is quenchable, a conclusion directly supported by our experimental synthesis (**Figure 3**). Conversely, the "Slow heating" pathway provides sufficient time for growth of α-Li$_3$PS$_4$ out of the nanoscale regime and into the stability fields of β- and eventually γ-Li$_3$PS$_4$ upon cooling, explaining why the bulk, high-temperature α-phase cannot be quenched.

The peculiar kinetic inhibition at 380 °C, where no crystalline phases appear except weak diffraction peaks at short periods of time (**Figure 2c**), can also be anticipated from the driving forces and nucleation kinetics. At this temperature, the free energy of the parent amorphous phase is very nearly thermoneutral ($\Delta G_{bulk} \sim 0$) with all the nanoscale Li$_3$PS$_4$ polymorphs. A small driving force results in a larger nucleation barrier for all phases; for example, the critical nucleation radius of α-Li$_3$PS$_4$ is very large, at 5.6 nm at 380 °C (**Figure S3c**); whereas the measured size of observed crystalline Li$_3$PS$_4$ in the amorphous matrix at 380 °C is only 5-20 nm from TEM. This explains why the amorphous phase can be more persistent within this very specific temperature window.

**Outlook**

Here we rationalized the nanoscale stabilization of quenchable metastable α-Li$_3$PS$_4$ by clarifying the underlying mechanisms that enables its synthesizaiblity. By integrating computational thermodynamics with high time-resolution experimental kinetic measurements, we constructed a Temperature-Time-Transformation diagram that quantitatively captures how rapid thermal processing promotes the remnant metastability of α-Li$_3$PS$_4$.[33,34] First-principles calculations showed that the low surface energy of the α phase thermodynamically stabilizes it at the nanoscale, a state achieved and maintained by rapid heating at relatively low temperatures that kinetically favors its nucleation while suppressing crystal growth. Validated by in-situ synchrotron XRD, this quantitative thermodynamic and kinetic picture provides a unified understanding of its nucleation and transformation pathways. This integrated methodology not only resolves the puzzle of α-Li$_3$PS$_4$, but also establishes a versatile platform applicable to other functional battery materials and glass-ceramics, paving the way for targeted synthesis of novel metastable phases.

Looking forward, the success of this strategy hinged on the synergistic combination of experimental and computational data, in order to deliver essential quantities that are inaccessible from either theory or experiment alone. While the energy of crystalline phases is well-established computationally, determining the temperature-dependent energy of an amorphous phase from first principles is notoriously difficult due to its non-periodic nature. To bridge this gap, here we used experimental differential scanning calorimetry (DSC) data to reference the bulk energy and heat capacity of the amorphous phase against the DFT-computed free energy of a crystalline phase. Similarly, when quantifying the kinetics, DFT-calculated bulk and surface energies can be used to estimate nucleation barriers, but the crystal growth rate is not easily accessible through computation alone. Therefore, integration of theory predictions with experimental high-time resolution measurements, such as diffraction or electron microscopy,[7,35-37] was essential for a holistic understanding of these kinetic factors.

**Methods**:

**Computational Methods**

Density Functional Theory (DFT) calculations were performed for atomistic structure calculations using Quantum Espresso (QE).[38,39] The revised regularized strongly constrained and appropriately normed ($r^2$SCAN) meta-Generalized Gradient Approximation (meta-GGA)[22] functional with revised Van der Waals 10 (rVV10) nonlocal correlation functional,[23] and norm-conserving pseudopotentials by Y. Yao and Y. Kanai[40] were applied to all DFT calculations of crystalline phases. The energy cutoff of wavefunction is set to 170 Ry and kinetic energy cutoff for charge density is set to 680 Ry from convergence test (**Figure S1**). The convergence threshold of the total energy, force, and stress is set to $10^{-4}$ eV/atom, $3 \times 10^{-3}$ eV/Å, and 0.5 kbar for structure relaxations, respectively. The k-points were optimally adapted from Pymatgen.[41]

Surface energies of the crystalline $Li_3PS_4$ were calculated from slab structures, generated by cleaving bulk phases using the Pymatgen package.[21] The maximum Miller indices was set to 3 and minimal slab size and vacuum length normal to the surface was set to 10-20 and 20 Å, respectively. The Wulff shape was also calculated using Pymatgen, and the weighted surface energy and shape factor from calculated surface energies was calculated by the γ-plot construction (**Table S1, S2 and S3**, **Figure S2**).[25] **Table S4** summarize all calculated parameters.

The vibrational free energy of three crystalline polymorphs was calculated from quasi-harmonic approximation. The ratio of unit cell volume was varied from 0.97 to 1.03 keeping the relaxed fractional atomic coordinates from structure relaxation using DFT. The atomic position displacement per each unit cell with varied volumes was generated with symmetry-adapted displacements for α-$Li_3PS_4$ and β-$Li_3PS_4$, and with 100 random displacements of γ-$Li_3PS_4$ on $2 \times 2 \times 2$ supercell. The total atomic energy of all generated structures was calculated using QE. The calculated energy with respect to the volume was fitted in the Rose-Vinet equation of states[42] and then obtained the lowest energy. The displaced structure generation and free energy calculations were performed using Phonopy[28,29] and with Alamode for random displacement.[30,31] The k-mesh parameters for phonon calculations were set to 8-12 times larger than the k-points for DFT calculations.

The configurational entropy contribution by Li disorder of α-$Li_3PS_4$ and β-$Li_3PS_4$ was also considered to total free energy. We randomly sampled 100 α-$Li_3PS_4$ and all 70 possible configurations of β-$Li_3PS_4$ by placing the Li to one of partial sites but not overlapping the same configuration of sampled ones keeping the original stoichiometry of $Li_3PS_4$ from the experimentally defined crystal structures and atomic positions with partial occupation sites[19] and calculated using QE (Supplementary Data). The configurational entropy was calculated from Boltzmann entropy formula (Supplementary Information). To define the contribution of partition function of configurations, the optimization of the energy contribution to statistical distribution was performed from Wang-Landau algorithm.[32] The energy was sampled from Monte Carlo up to 100,000 steps in maximum and repeated the calculation with various temperatures.

The free energy of the amorphous phase was calculated from the experimental differential scanning calorimetry (DSC) reference data.[3] The 3 mg $Li_3PS_4$ glass powder was placed in an Al pan under a dry Ar atmosphere and performed DSC measurements heating to 500 °C and then cooled to 25 °C at a scanning rate of 10 °C/min. The DSC during the heating cycle observed an exothermic peak at 230 °C showing phase transition from amorphous to β-$Li_3PS_4$. The enthalpy difference between amorphous and β-$Li_3PS_4$ was calculated by the integration of analyzed thermal peaks from the corrected baseline. The entropy of amorphous $Li_3PS_4$ were calculated from the integrated heat capacity into the entropy change equation at constant pressure (Supplementary Information). The total free energy of amorphous $Li_3PS_4$ is calculated from 25 to 230 °C before phase transition. Calculated free energy data points of amorphous and the melting temperature of crystalline $Li_3PS_4$ at 700 °C were fit to a power function for the entire temperature range (**Figure S3d**).

**Experimental Methods**

All the starting materials and products were handled without exposure to air. The $Li_2S$-$P_2S_5$ samples were first prepared by hand milling with a mortar and pestle. Subsequent ball-milling was performed with a zirconia pot (45 ml) and a zirconia ball (6 mm diameter). In-situ synchrotron X-ray diffraction was performed at the BL13XU beamline in the SPring-8 with the approvals of 2023B1669, 2023B1942 and 2025B1753. The samples were sealed in a quartz capillary with a diameter of 0.3 mm. The samples were heated by a nitrogen flow, and diffraction patterns were recorded by LAMBDA 750 K detectors.[43] The wavelength was 0.0354 nm or 0.0249nm.

In-situ SXRD was performed in two ways. One was at the constant heating rate of 60 °C/min, and the other was isothermal experiment. For the isothermal experiment, a nitrogen gas flow for heating the capillary samples was preheated at 220, 237, 260, 300, 340, 360, 367, 380, 440, 500, and 547 °C without the samples. The diffraction measurements were collected every 0.5 second, and the samples were inserted in a preheated nitrogen flow. The samples were kept during the isothermal measurements. We defined the time of 0.5 seconds when the diffraction data was detected. We measured the change in lattice parameters of silver particles in the vacuumed capillary in the same setting of isothermal reactions. This measurement showed that the lattice parameters are almost constant after 2.0 seconds. Thus, this measurement is the rapid heating within 2.0 seconds and subsequent isothermal heating after 2.0 seconds (**Figure S5**) Rietveld analysis was performed to quantify the phase fraction using TOPAS software.[44] We normalized the mass fraction as 100 % using the maximum sum scale factors of the alpha and beta phases in each isothermal measurement, and a time-temperature-transformation diagram was constructed.

Guided by the TTT diagram, glass samples were crystallized in a short time. A few tens of milligrams of glass sample produced by ball milling were put on the hot plate and pressed with a plastering spatula for 10 seconds and collected. The product was characterized using SXRD and laboratory XRD with CuKa radiation by Miniflux600 (Rigaku). Impedance spectra were measured by SP-200 (BioLogic) to estimate its ion conductivity (**Figure S6**).

In-situ TEM observation was conducted by JEM-2100Plus (JEOL Co. Ltd) with an accelerated voltage of 200kV. In this in-situ experiment, 4 Electrodes Vacuum Transfer Holder (Mel Build Co. Ltd) was used to prevent air exposure during sample transfer from the glovebox to the TEM. In addition, a high-performance heating micro-electromechanical systems (MEMS) chip was used to control the rate of temperature change.

**Acknowledgements**
A. Miura and W. Baek acknowledge T. Kimura for providing DSC data of $Li_3PS_4$ . This study was partially supported by JST PRESTO JPMJPR21Q8, JST Gtex JPMJGX23S5, JSPS KAKENHI 20KK0124 and 24K21775, JST ASPIRE JPMJAP2419 and JSPS Program for Forming Japan's Peak Research Universities (J-PEAKS) Grant Number JPJS00420230001. The work by W. Baek and W. Sun was supported by the US Department of Energy (DOE), Office of Science, Basic Energy Sciences (Award No. DE-SC0021130). W. Baek and W. Sun acknowledge Texas Advanced Computing Center at the University of Texas at Austin and Rosen Center for Advanced Computing at Purdue University, which uses Grant Award TG-MAT240013 from the Advanced Cyberinfrastructure Coordination Ecosystem: Services & Support (US National Science Foundation Award number #2138259, #2138286, #2138307, #2137603 and #2138296), National Energy Research Scientific Computing Center at Lawrence Berkeley National Laboratory, which is supported by the Office of Science of the DOE (Contract No. DE-AC02-05CH11231 using Award No. BES-ERCAP0033722), and Information and Technology Service at the University of Michigan, for providing high-performance computing resources that have contributed to the research results.

## Author contributions

Conceptualization: A. M., W. B. and W. S., Methodology: S. Kobayashi, S. Kawaguchi. and W. B., Investigation: A. M., W. B. Y. F., Y. A., Y. M., C. M., D. J. Visualization: A. M., M. H., and W. B. Funding acquisition: A. M., A. H. and W. S. Resources: K. T., A. S. and S. M. Supervision: A. M. and W. S., Writing – original draft: A. M., and W. B., Writing – review and editing: A. M., W. B., and W. S.

**Competing interests**: We declare no competing interests.

**Data and materials availability**: All data are available in the manuscript or the Supplementary Information and Data.

## References

1  England, W. A., Goodenough, J. B. & Wiseman, P. J. Ion-exchange reactions of mixed oxides. *J. Solid State Chem.* **49**, 289-299 (1983). https://doi.org:https://doi.org/10.1016/S0022-4596(83)80006-1

2  Sasahara, Y. *et al.* Mechanochemical Synthesis of Perovskite Oxyhydrides: Insights from Shear Modulus. *Journal of the American Chemical Society* **146**, 11694-11701 (2024). https://doi.org:10.1021/jacs.3c14087

3  Kimura, T. *et al.* Stabilizing High-Temperature $\alpha$-Li$_3$PS$_4$ by Rapidly Heating the Glass. *Journal of the American Chemical Society* **145**, 14466-14474 (2023). https://doi.org:10.1021/jacs.3c03827

4  Kurdjumow, G. & Sachs, G. Über den Mechanismus der Stahlhärtung. *Zeitschrift für Physik* **64**, 325-343 (1930). https://doi.org:10.1007/BF01397346

5  Sun, W., Powell-Palm, M. & Chen, J. *The geometry of high-dimensional phase diagrams: I. Generalized Gibbs' Phase Rule* (2024).

6  Bianchini, M. *et al.* The interplay between thermodynamics and kinetics in the solid-state synthesis of layered oxides. *Nat Mater* (2020). https://doi.org:10.1038/s41563-020-0688-6

7  Miura, A. *et al.* Observing and Modeling the Sequential Pairwise Reactions that Drive Solid-State Ceramic Synthesis. *Advanced Materials* **33**, e2100312 (2021). https://doi.org:10.1002/adma.202100312

8  Chen, B. R. *et al.* Understanding crystallization pathways leading to manganese oxide polymorph formation. *Nat Commun* **9**, 2553 (2018). https://doi.org:10.1038/s41467-

018-04917-y

9    Sun, W., Kitchaev, D. A., Kramer, D. & Ceder, G. Non-equilibrium crystallization pathways of manganese oxides in aqueous solution. *Nat Commun* **10**, 573 (2019). https://doi.org/10.1038/s41467-019-08494-6

10   Baek, W., Das, S., Tan, S., Gavini, V. & Sun, W. Quasicrystal stability and nucleation kinetics from density functional theory. *Nature Physics* **21**, 980-987 (2025). https://doi.org/10.1038/s41567-025-02925-6

11   Rom, C. L. *et al.* Mechanistically Guided Materials Chemistry: Synthesis of Ternary Nitrides, CaZrN$_2$ and CaHfN$_2$. *Journal of the American Chemical Society* **146**, 4001-4012 (2024). https://doi.org:10.1021/jacs.3c12114

12   Chen, J., Powell-Palm, M. & Sun, W. *The geometry of high-dimensional phase diagrams: II. The duality between closed and open chemical systems.* (2024).

13   Davenport, E. S. & Bain, E. C. Transformation of austenite at constant subcritical temperatures. *Metallurgical Transactions* **1**, 3503-3530 (1970). https://doi.org:10.1007/BF03037892

14   Sasaki, S. *et al.* Crystallization and vitrification of electrons in a glass-forming charge liquid. *Science* **357**, 1381-1385 (2017). https://doi.org:doi:10.1126/science.aal3120

15   Ariyoshi, K., Yuzawa, K. & Yamada, Y. Reaction Mechanism and Kinetic Analysis of the Solid-State Reaction to Synthesize Single-Phase Li$_2$Co2O$_4$ Spinel. *The Journal of Physical Chemistry C* **124**, 8170-8177 (2020). https://doi.org:10.1021/acs.jpcc.0c01115

16   Zhu, Y., Chon, M., Thompson, C. V. & Rupp, J. L. M. Time-Temperature-Transformation (TTT) Diagram of Battery-Grade Li-Garnet Electrolytes for Low-Temperature Sustainable Synthesis. *Angew Chem Int Ed Engl* **62**, e202304581 (2023). https://doi.org:10.1002/anie.202304581

17   Hu, D. *et al.* The RAPTR furnace: a rapid heating and cooling sample furnace for in situ X-ray scattering studies of temperature-induced reactions. *Journal of Applied Crystallography* **57**, 88-93 (2024). https://doi.org:10.1107/s1600576723011020

18   Yamamoto, T. *et al.* Emergence of Dynamically-Disordered Phases During Fast Oxygen Deintercalation Reaction of Layered Perovskite. *Adv Sci (Weinh)* **10**, e2301876 (2023). https://doi.org:10.1002/advs.202301876

19   Homma, K. *et al.* Crystal structure and phase transitions of the lithium ionic conductor Li$_3$PS$_4$. *Solid State Ionics* **182**, 53-58 (2011). https://doi.org:10.1016/j.ssi.2010.10.001

20   Kaup, K., Zhou, L., Huq, A. & Nazar, L. F. Impact of the Li substructure on the diffusion pathways in alpha and beta Li$_3$PS$_4$: an in situ high temperature neutron diffraction study. *Journal of Materials Chemistry A* (2020). https://doi.org:10.1039/d0ta02805c

21   Sun, W. & Ceder, G. Efficient creation and convergence of surface slabs. *Surface Science*

**617**, 53-59 (2013). https://doi.org:10.1016/j.susc.2013.05.016

22    Furness, J. W., Kaplan, A. D., Ning, J., Perdew, J. P. & Sun, J. Accurate and Numerically Efficient r2SCAN Meta-Generalized Gradient Approximation. *The Journal of Physical Chemistry Letters* **11**, 8208-8215 (2020). https://doi.org:10.1021/acs.jpclett.0c02405

23    Peng, H., Yang, Z.-H., Perdew, J. P. & Sun, J. Versatile van der Waals Density Functional Based on a Meta-Generalized Gradient Approximation. *Physical Review X* **6** (2016). https://doi.org:10.1103/PhysRevX.6.041005

24    Tran, R. *et al.* Surface energies of elemental crystals. *Scientific Data* **3**, 160080 (2016). https://doi.org:10.1038/sdata.2016.80

25    Herring, C. Some Theorems on the Free Energies of Crystal Surfaces. *Physical Review* **82**, 87-93 (1951). https://doi.org:10.1103/PhysRev.82.87

26    Navrotsky, A. Energetic clues to pathways to biomineralization: Precursors, clusters, and nanoparticles. *Proceedings of the National Academy of Sciences* **101**, 12096-12101 (2004). https://doi.org:doi:10.1073/pnas.0404778101

27    Janek, J., Martin, M. & Becker, K. D. Physical chemistry of solids--the science behind materials engineering. *Physical chemistry chemical physics : PCCP* **11**, 3010 (2009). https://doi.org:10.1039/b905911n

28    Togo, A. First-principles Phonon Calculations with Phonopy and Phono3py. *Journal of the Physical Society of Japan* **92**, 012001 (2022). https://doi.org:10.7566/JPSJ.92.012001

29    Togo, A., Chaput, L., Tadano, T. & Tanaka, I. Implementation strategies in phonopy and phono3py. *Journal of Physics: Condensed Matter* **35**, 353001 (2023). https://doi.org:10.1088/1361-648X/acd831

30    Tadano, T., Gohda, Y. & Tsuneyuki, S. Anharmonic force constants extracted from first-principles molecular dynamics: applications to heat transfer simulations. *Journal of Physics: Condensed Matter* **26**, 225402 (2014). https://doi.org:10.1088/0953-8984/26/22/225402

31    Tadano, T. & Tsuneyuki, S. First-Principles Lattice Dynamics Method for Strongly Anharmonic Crystals. *Journal of the Physical Society of Japan* **87**, 041015 (2018). https://doi.org:10.7566/JPSJ.87.041015

32    Wang, F. & Landau, D. P. Efficient, multiple-range random walk algorithm to calculate the density of states. *Phys Rev Lett* **86**, 2050-2053 (2001). https://doi.org:10.1103/PhysRevLett.86.2050

33    Sun, W. *et al.* The thermodynamic scale of inorganic crystalline metastability. *Science advances* **2**, e1600225-e1600225 (2016). https://doi.org:10.1126/sciadv.1600225

34    Ito, H. *et al.* Stability and Metastability of $Li_3YCl_6$ and $Li_3HoCl_6$. *Bulletin of the Chemical*

*Society of Japan* **96**, 1262-1268 (2023). https://doi.org:10.1246/bcsj.20230132

35  Kim, J. *et al.* Dissolution enables dolomite crystal growth near ambient conditions. *Science* **382**, 915-920 (2023). https://doi.org:10.1126/science.adi3690

36  Sakakibara, M., Hanaya, M., Nakamuro, T. & Nakamura, E. Nondeterministic dynamics in the $\eta$-to-$\theta$ phase transition of alumina nanoparticles. *Science* **387**, 522-527 (2025). https://doi.org:doi:10.1126/science.adr8891

37  Hisasue, R. *et al.* Kinetic effects of ball-milling precursors on the synthesis pathway of KSbF4. *Ceramics International* **51**, 33653-33660 (2025). https://doi.org:10.1016/j.ceramint.2025.05.096

38  Giannozzi, P. *et al.* QUANTUM ESPRESSO: a modular and open-source software project for quantum simulations of materials. *Journal of physics. Condensed matter : an Institute of Physics journal* **21**, 395502 (2009). https://doi.org:10.1088/0953-8984/21/39/395502

39  Giannozzi, P. *et al.* Advanced capabilities for materials modelling with Quantum ESPRESSO. *Journal of physics. Condensed matter : an Institute of Physics journal* **29**, 465901 (2017). https://doi.org:10.1088/1361-648X/aa8f79

40  Yao, Y. & Kanai, Y. Plane-wave pseudopotential implementation and performance of SCAN meta-GGA exchange-correlation functional for extended systems. *The Journal of Chemical Physics* **146** (2017). https://doi.org:10.1063/1.4984939

41  Jain, A. *et al.* A high-throughput infrastructure for density functional theory calculations. *Computational Materials Science* **50**, 2295-2310 (2011). https://doi.org:https://doi.org/10.1016/j.commatsci.2011.02.023

42  Vinet, P., Smith, J. R., Ferrante, J. & Rose, J. H. Temperature effects on the universal equation of state of solids. *Physical Review B* **35**, 1945-1953 (1987). https://doi.org:10.1103/PhysRevB.35.1945

43  Kawaguchi, S. *et al.* High-throughput and high-resolution powder X-ray diffractometer consisting of six sets of 2D CdTe detectors with variable sample-to-detector distance and innovative automation system. *Journal of Synchrotron Radiation* **31**, 955-967 (2024). https://doi.org:doi:10.1107/S1600577524003539

44  Coelho, A. A. TOPAS and TOPAS-Academic: an optimization program integrating computer algebra and crystallographic objects written in C++. *Journal of Applied Crystallography* **51**, 210-218 (2018). https://doi.org:10.1107/s1600576718000183

# Supporting Information: Time-Temperature-Transformation (TTT) Diagrams to rationalize the nucleation and quenchability of metastable α-Li₃PS₄

Akira Miura,[a,†,*] Woohyeon Baek,[b,†] Yuta Fujii,[a] Kiyoharu Tadanaga,[a] Rana Hossain,[a] Yamashita Aichi,[c] Yoshikazu Mizuguchi,[c] Chikako Moriyoshi,[d] Shintaro Kobayashi,[e] Shogo Kawaguchi,[e] Ding Jiong,[f] Shigeo Mori,[f] Atsushi Sakuda,[g] Akitoshi Hayashi,[g,h] Wenhao Sun[b,*]

[a] Faculty of Engineering, Hokkaido University, Kita 13, Nishi 8, Sapporo, 060-8628, Japan

[b] Department of Materials Science and Engineering, University of Michigan, Ann Arbor, MI, 48109, USA

[c] Department of Physics, Tokyo Metropolitan University, Hachioji, 192-0397, Japan

[d] Graduate School of Advanced Science and Engineering, Hiroshima University, 1-3-1, Kagamiyama, Higashihiroshima 739-8526, Japan

[e] Japan Synchrotron Radiation Research Institute, Sayo, Hyogo, 679-5198, Japan

[f] Department of Materials Science, Graduate School of Engineering, Osaka Metropolitan University, 1-1 Gakuen-cho, Naka-ku, Sakai, Osaka, 599-8531, Japan

[g] Department of Applied Chemistry, Graduate School of Engineering, Osaka Metropolitan University, 1-1 Gakuen-cho, Naka-ku, Sakai, Osaka, 599-8531, Japan

[h] Institute for Materials Research, Tohoku University, 2-1-1 Katahira, Aoba-ku, Sendai, Miyagi 980-8577, Japan

† contributed equally    Email: amiura@eng.hokudai.ac.jp (AM), whsun@umich.edu (WS)

**Table S1.** Calculated surface energies of crystalline α-Li$_3$PS$_4$ polymorphs from slab structures.

| plane | γ (eV/Å²) | γ (J/m²) | plane | γ (eV/Å²) | γ (J/m²) |
|---|---|---|---|---|---|
| (0, 0, 1) | 0.0665 | 1.0653 | (2, 0, 1) | 0.0356 | 0.5697 |
| (0, 1, 0) | 0.0736 | 1.1798 | (2, 0, 3) | 0.0324 | 0.5192 |
| (0, 1, 1) | 0.0582 | 0.9317 | (2, 1, 0) | 0.0493 | 0.7894 |
| (0, 1, 2) | 0.0530 | 0.8495 | (2, 1, 1) | 0.0474 | 0.7599 |
| (0, 1, 3) | 0.0473 | 0.7583 | (2, 1, 2) | 0.0553 | 0.8864 |
| (0, 2, 1) | 0.0583 | 0.9340 | (2, 1, 3) | 0.0342 | 0.5473 |
| (0, 2, 3) | 0.0478 | 0.7664 | (2, 2, 1) | 0.0412 | 0.6594 |
| (0, 3, 1) | 0.0596 | 0.9546 | (2, 2, 3) | 0.0473 | 0.7575 |
| (0, 3, 2) | 0.0535 | 0.8567 | (2, 3, 0) | 0.0453 | 0.7264 |
| (1, 0, 0) | 0.0692 | 1.1089 | (2, 3, 1) | 0.0537 | 0.8611 |
| (1, 0, 1) | 0.0271 | 0.4344 | (2, 3, 2) | 0.0616 | 0.9867 |
| (1, 0, 2) | 0.0400 | 0.6408 | (2, 3, 3) | 0.0516 | 0.8270 |
| (1, 0, 3) | 0.0415 | 0.6647 | (3, 0, 1) | 0.0406 | 0.6502 |
| (1, 1, 0) | 0.0529 | 0.8474 | (3, 0, 2) | 0.0355 | 0.5687 |
| (1, 1, 1) | 0.0520 | 0.8334 | (3, 1, 0) | 0.0467 | 0.7475 |
| (1, 1, 2) | 0.0373 | 0.5969 | (3, 1, 1) | 0.0352 | 0.5641 |
| (1, 1, 3) | 0.0392 | 0.6285 | (3, 1, 2) | 0.0311 | 0.4985 |
| (1, 2, 0) | 0.0410 | 0.6563 | (3, 1, 3) | 0.0247 | 0.3956 |
| (1, 2, 1) | 0.0569 | 0.9112 | (3, 2, 0) | 0.0365 | 0.5846 |
| (1, 2, 2) | 0.0498 | 0.7973 | (3, 2, 1) | 0.0463 | 0.7410 |
| (1, 2, 3) | 0.0508 | 0.8133 | (3, 2, 2) | 0.0481 | 0.7706 |
| (1, 3, 0) | 0.0473 | 0.7576 | (3, 2, 3) | 0.0558 | 0.8937 |
| (1, 3, 1) | 0.0494 | 0.7917 | (3, 3, 1) | 0.0313 | 0.5007 |
| (1, 3, 2) | 0.0595 | 0.9540 | (3, 3, 2) | 0.0629 | 1.0085 |
| (1, 3, 3) | 0.0516 | 0.8267 | | | |

**Table S2.** Calculated surface energies of crystalline β-Li$_3$PS$_4$ polymorphs from slab structures.

| plane | γ (eV/Å²) | γ (J/m²) | plane | γ (eV/Å²) | γ (J/m²) |
|---|---|---|---|---|---|
| (0, 0, 1) | 0.0660 | 1.0575 | (2, 0, 1) | 0.0550 | 0.8805 |
| (0, 1, 0) | 0.0521 | 0.8343 | (2, 0, 3) | 0.0653 | 1.0455 |
| (0, 1, 1) | 0.0457 | 0.7320 | (2, 1, 0) | 0.0429 | 0.6879 |
| (0, 1, 2) | 0.0576 | 0.9222 | (2, 1, 1) | 0.0368 | 0.5904 |
| (0, 1, 3) | 0.0814 | 1.3044 | (2, 1, 2) | 0.0429 | 0.6879 |
| (0, 2, 1) | 0.0589 | 0.9432 | (2, 1, 3) | 0.0453 | 0.7251 |
| (0, 2, 3) | 0.0711 | 1.1399 | (2, 2, 1) | 0.0455 | 0.7290 |
| (0, 3, 1) | 0.0659 | 1.0566 | (2, 2, 3) | 0.0358 | 0.5744 |
| (0, 3, 2) | 0.0500 | 0.8012 | (2, 3, 0) | 0.0454 | 0.7266 |
| (1, 0, 0) | 0.0474 | 0.7590 | (2, 3, 1) | 0.0433 | 0.6941 |
| (1, 0, 1) | 0.0366 | 0.5862 | (2, 3, 2) | 0.0474 | 0.7596 |
| (1, 0, 2) | 0.0573 | 0.9186 | (2, 3, 3) | 0.0352 | 0.5641 |
| (1, 0, 3) | 0.0437 | 0.7006 | (3, 0, 1) | 0.0413 | 0.6617 |
| (1, 1, 0) | 0.0330 | 0.5292 | (3, 0, 2) | 0.0396 | 0.6349 |
| (1, 1, 1) | 0.0479 | 0.7674 | (3, 1, 0) | 0.0362 | 0.5793 |
| (1, 1, 2) | 0.0302 | 0.4838 | (3, 1, 1) | 0.0371 | 0.5950 |
| (1, 1, 3) | 0.0585 | 0.9374 | (3, 1, 2) | 0.0447 | 0.7166 |
| (1, 2, 0) | 0.0446 | 0.7148 | (3, 1, 3) | 0.0398 | 0.6379 |
| (1, 2, 1) | 0.0330 | 0.5285 | (3, 2, 0) | 0.0395 | 0.6322 |
| (1, 2, 2) | 0.0336 | 0.5388 | (3, 2, 1) | 0.0381 | 0.6108 |
| (1, 2, 3) | 0.0350 | 0.5615 | (3, 2, 2) | 0.0377 | 0.6035 |
| (1, 3, 0) | 0.0447 | 0.7160 | (3, 2, 3) | 0.0338 | 0.5415 |
| (1, 3, 1) | 0.0440 | 0.7050 | (3, 3, 1) | 0.0405 | 0.6482 |
| (1, 3, 2) | 0.0424 | 0.6798 | (3, 3, 2) | 0.0382 | 0.6121 |
| (1, 3, 3) | 0.0335 | 0.5374 | | | |

**Table S3.** Calculated surface energies of crystalline γ-Li₃PS₄ polymorphs from slab structures.

| plane | γ (eV/Å²) | γ (J/m²) | plane | γ (eV/Å²) | γ (J/m²) |
|---|---|---|---|---|---|
| (0, 0, 1) | 0.0632 | 1.0121 | (2, 0, 1) | 0.0597 | 0.9570 |
| (0, 1, 0) | 0.0508 | 0.8141 | (2, 0, 3) | 0.0517 | 0.8281 |
| (0, 1, 1) | 0.0759 | 1.2155 | (2, 1, 0) | 0.0712 | 1.1406 |
| (0, 1, 2) | 0.0486 | 0.7781 | (2, 1, 1) | 0.0369 | 0.5906 |
| (0, 1, 3) | 0.0497 | 0.7964 | (2, 1, 2) | 0.0454 | 0.7282 |
| (0, 2, 1) | 0.0467 | 0.7489 | (2, 1, 3) | 0.0538 | 0.8626 |
| (0, 2, 3) | 0.0585 | 0.9368 | (2, 2, 1) | 0.0568 | 0.9104 |
| (0, 3, 1) | 0.0590 | 0.9458 | (2, 2, 3) | 0.0567 | 0.9078 |
| (0, 3, 2) | 0.0562 | 0.9004 | (2, 3, 0) | 0.0616 | 0.9874 |
| (1, 0, 0) | 0.0679 | 1.0876 | (2, 3, 1) | 0.0475 | 0.7617 |
| (1, 0, 1) | 0.1086 | 1.7397 | (2, 3, 2) | 0.0494 | 0.7919 |
| (1, 0, 2) | 0.0595 | 0.9527 | (2, 3, 3) | 0.0527 | 0.8442 |
| (1, 0, 3) | 0.0625 | 1.0012 | (3, 0, 1) | 0.0627 | 1.0043 |
| (1, 1, 0) | 0.0793 | 1.2700 | (3, 0, 2) | 0.0624 | 1.0005 |
| (1, 1, 1) | 0.0423 | 0.6782 | (3, 1, 0) | 0.0580 | 0.9298 |
| (1, 1, 2) | 0.0599 | 0.9590 | (3, 1, 1) | 0.0408 | 0.6539 |
| (1, 1, 3) | 0.0522 | 0.8358 | (3, 1, 2) | 0.0543 | 0.8700 |
| (1, 2, 0) | 0.0546 | 0.8751 | (3, 1, 3) | 0.0684 | 1.0952 |
| (1, 2, 1) | 0.0436 | 0.6982 | (3, 2, 0) | 0.0577 | 0.9242 |
| (1, 2, 2) | 0.0612 | 0.9812 | (3, 2, 1) | 0.0324 | 0.5192 |
| (1, 2, 3) | 0.0597 | 0.9567 | (3, 2, 2) | 0.0505 | 0.8095 |
| (1, 3, 0) | 0.0442 | 0.7086 | (3, 2, 3) | 0.0489 | 0.7833 |
| (1, 3, 1) | 0.0372 | 0.5966 | (3, 3, 1) | 0.0565 | 0.9059 |
| (1, 3, 2) | 0.0537 | 0.8597 | (3, 3, 2) | 0.0430 | 0.6894 |
| (1, 3, 3) | 0.0634 | 1.0157 | | | |

**Table S4.** Calculated parameters for the Wulff construction of crystalline Li$_3$PS$_4$ polymorphs from the periodic and slab calculations. The bulk formation energy is calculated with reference to bulk Li$_2$S and P$_2$S$_5$.

|  | α-Li$_3$PS$_4$ | β-Li$_3$PS$_4$ | γ-Li$_3$PS$_4$ |
|---:|:---:|:---:|:---:|
| Shape factor (η) | 5.5619 | 5.2678 | 5.2842 |
| Atomic density (ρ) (atom/Å$^3$) | 0.0468 | 0.0467 | 0.0486 |
| Weighted surface energy (γ) (eV/Å$^2$) | 0.0268 | 0.0314 | 0.0360 |
| Weighted surface energy (γ) (J/m$^2$) | 0.4294 | 0.5031 | 0.5768 |
| Bulk formation energy ($E_{bulk}$/N) (eV/atom) | -0.1147 | -0.1234 | -0.1395 |
| Bulk formation energy ($E_{bulk}$/N) (kJ/mol) | -11.0668 | -11.9063 | -13.4597 |

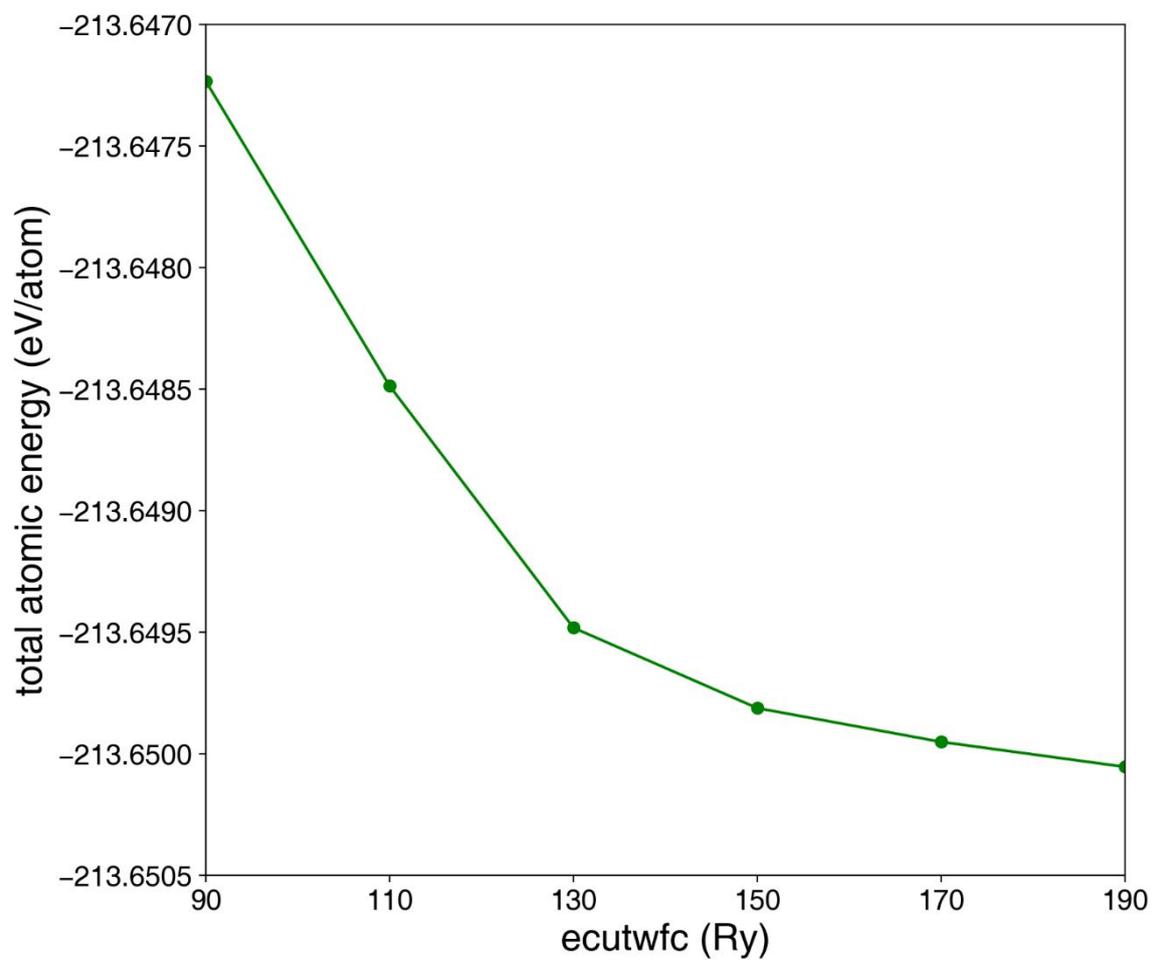

**Figure S1.** Total atomic energy of γ-Li$_3$PS$_4$ from QE and norm-conserving pseudopotentials by Y. Yao and Y. Kanai.

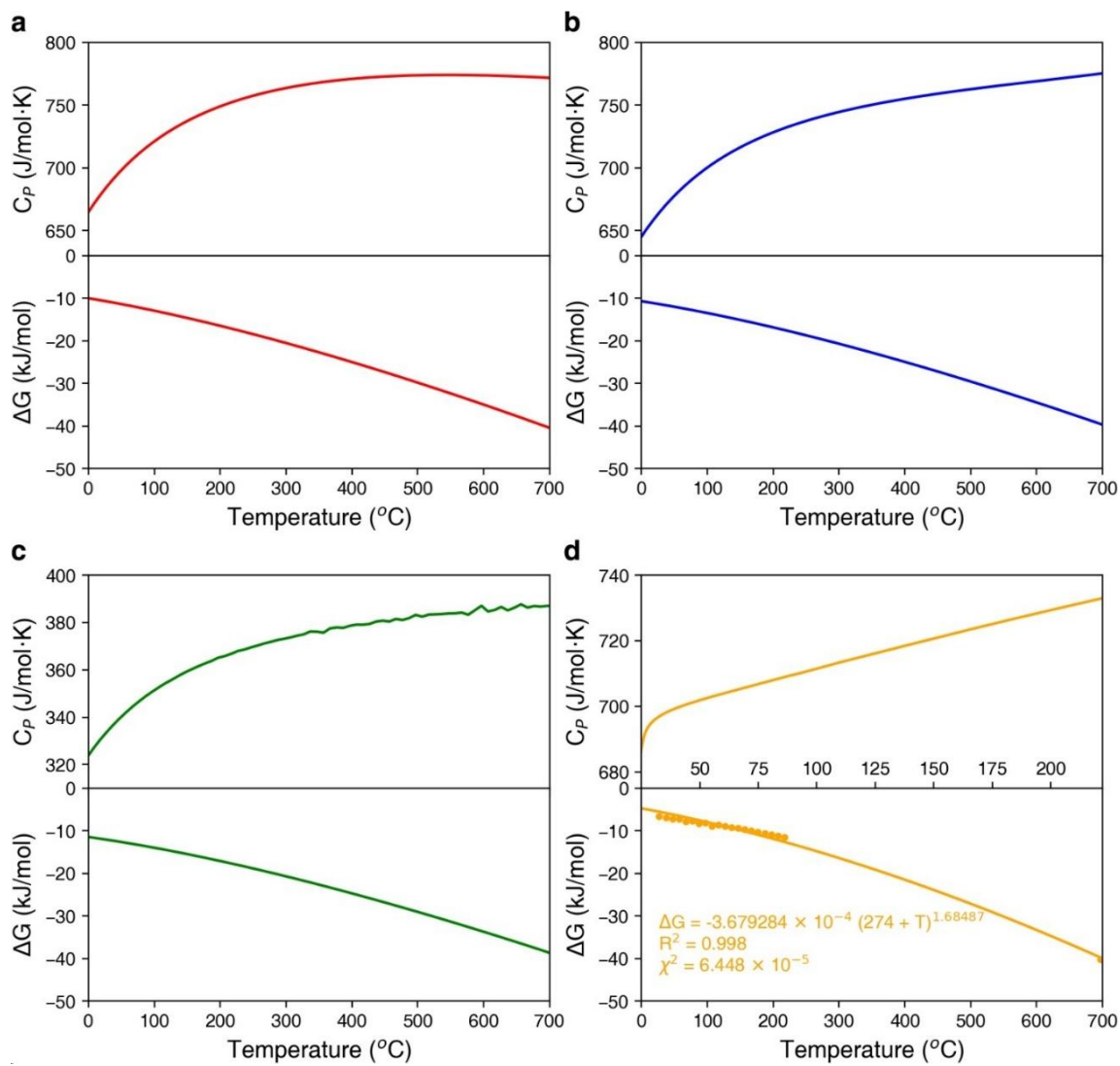

**Figure S2.** Heat capacity and free energy of $Li_3PS_4$ polymorphs: **a,** α-$Li_3PS_4$, **b,** β-$Li_3PS_4$, **c,** γ-$Li_3PS_4$, and **d,** amorphous $Li_3PS_4$. The crystalline polymorphs were calculated from DFT/Phonopy and amorphous from the experimental DSC reference data with respect to β-$Li_3PS_4$. The configurational entropy contribution of α-$Li_3PS_4$ and β-$Li_3PS_4$ is included in the free energy. The free energy is referenced to bulk $Li_2S$ and $P_2S_5$.

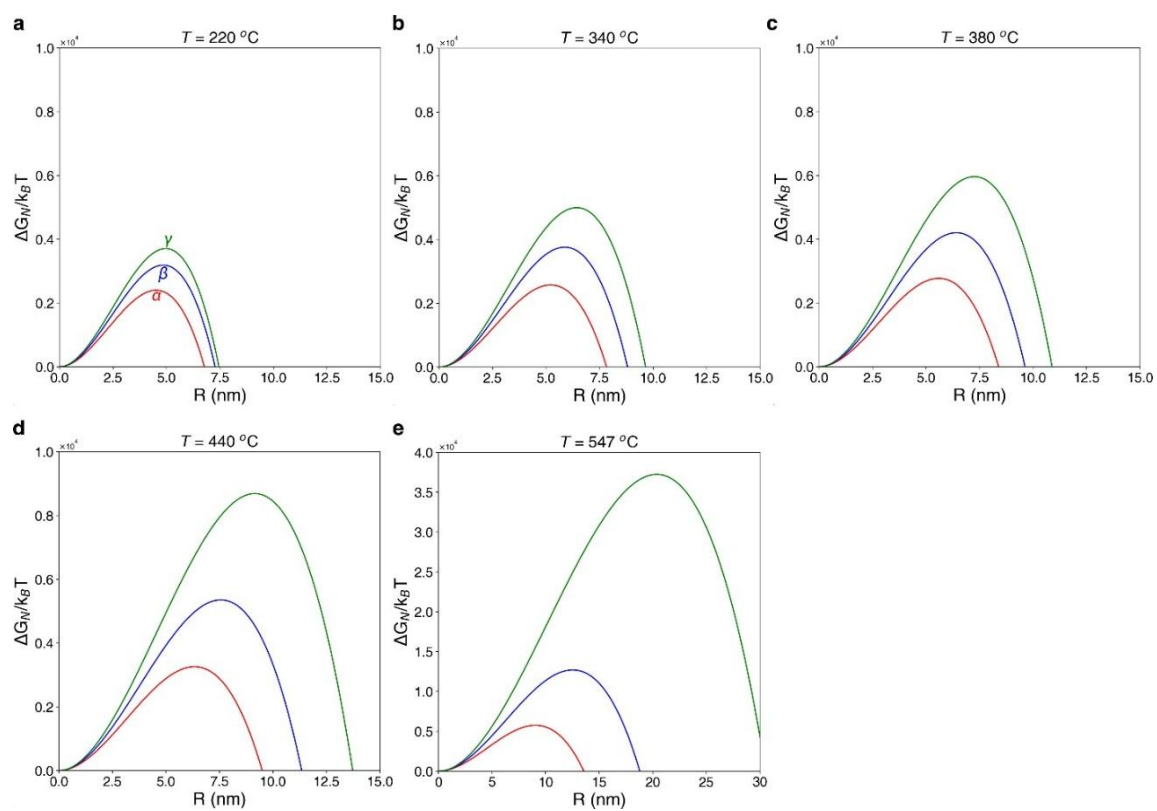

**Figure S3.** Free energy diagram of the homogeneous nucleation from the bulk amorphous matrix to crystalline Li$_3$PS$_4$ polymorphs at **a,** 220, **b,** 340, **c,** 380, **d,** 440, and **e,** 547 °C.

The Wulff construction of all crystalline phases is built by calculating the surface energy of cleaved faces.

$$\gamma = \frac{E_{slab} - N_{slab}\left(\frac{E_{bulk}}{N_{bulk}}\right)}{2A}$$

where $E_{slab}$ is the energy and $N_{slab}$ is the total number of atoms in a slab cell, and $E_{bulk}$ and $N_{bulk}$ are in bulk periodic cell.

From the calculated parameters, theoretical total formation energy at a nano-scale system can be calculated as a function of Area-to-Volume ratio, or equivalently, $1/R$.[1,2]

$$\frac{E_{NP}}{N} = \frac{E_{bulk}}{N} + \frac{\gamma \eta}{\rho} \cdot \frac{1}{R}$$

where $\eta$ is dimensionless anisotropic shape factor, $\rho$ is atomic density, and R is radius of nanoclusters.

The configurational entropy of Boltzmann entropy formula is

$$S = -k_B \sum_i P_i \ln P_i, \quad P_i = \frac{e^{-\frac{E_i}{k_B T}}}{Z}$$

where $k_B$ is Boltzmann constant and P is a probability of configurations. The partition function (Z) of configurations in Wang-Landau algorithm is[3]

$$Z = \sum_i g(E_i) e^{-\frac{E_i}{k_B T}}$$

where g(E) is the density of states of energy level.

The heat capacity ($C_P$), enthalpy change ($\Delta H$), and entropy change ($\Delta S$) from direct scanning calorimetry (DSC) are

$$C_P = \frac{\Delta Q}{\beta m}, \quad \Delta H = \frac{\int Q dt}{m}, \quad \Delta S = \int \frac{C_P}{T} dT$$

where Q is heat flow, β is heating rate, t is time, m is mass of a sample, and T is temperature.

The free energy of a nucleus ($\Delta G_N$) and at the critical nucleus ($\Delta G_C$) of the homogeneous nucleation are[2,4]

$$\Delta G_N = \rho R^3 \Delta G_{amorphous-solid} + \gamma \eta R^2, \quad \Delta G_C = \frac{4(\gamma \eta)^3}{27\left(\rho \Delta G_{amorphous-solid}\right)^2}$$

where γ is surface energy, η is dimensionless shape factor, and ρ is atomic density. The $\Delta G_{amorphous-solid}$ is the driving force calculated from the energy difference between amorphous and solid phase.

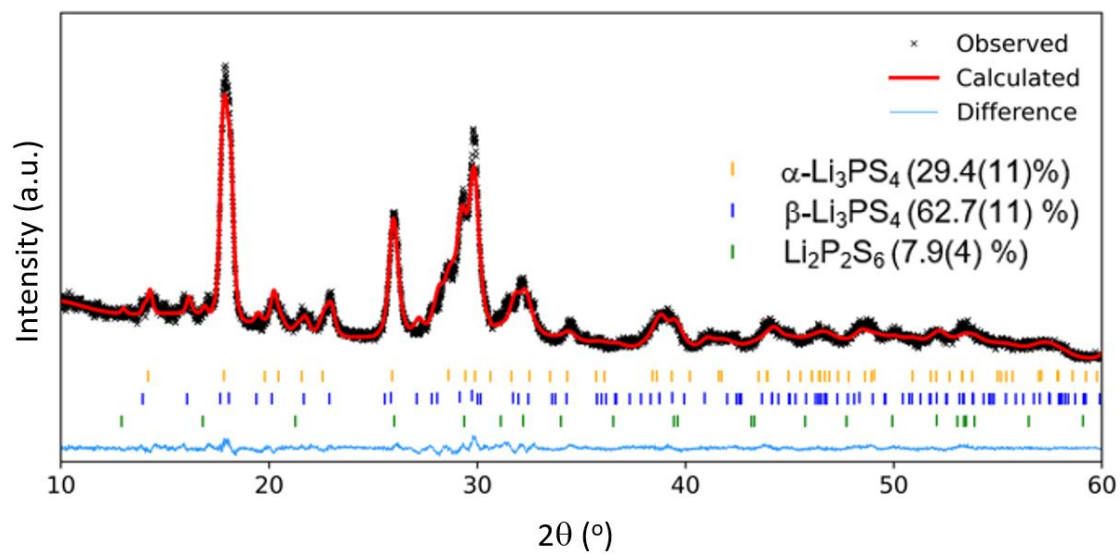

**Figure S4.** Rietveld refinement of Li$_3$PS$_4$ after fast heating at 350 °C for 10 seconds.

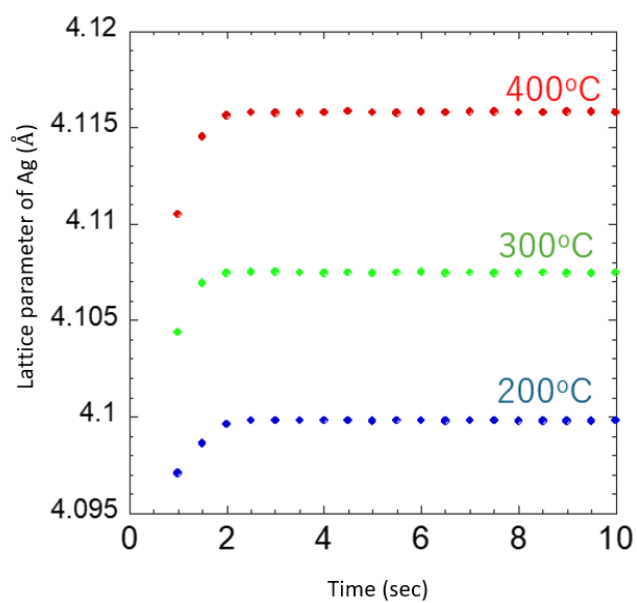

**Figure S5.** The time dependence of lattice parameters of silver metals with different isothermal heating at 200, 300, and 400 °C. Quick lattice expansion within 2 sec and subsequent constant lattice parameters indicate that the heating profile is rapid heating within 2 sec and subsequent isothermal heating.

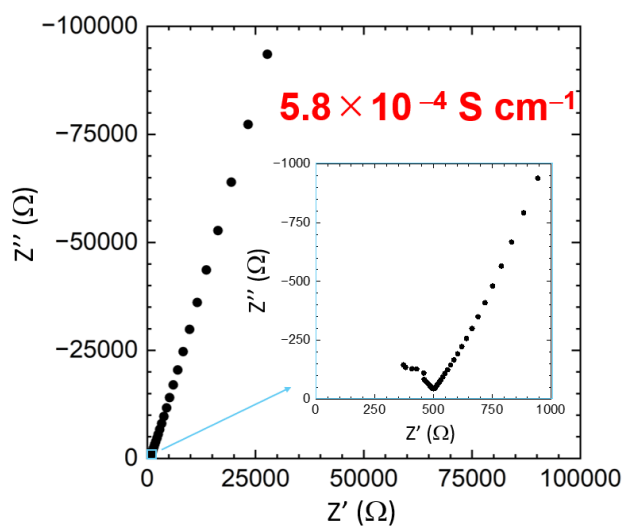

**Figure S6.** AC impedance measurement of Li$_3$PS$_4$ after fast heating at 350 °C for 10 seconds.

References

1. Sun, W., Kitchaev, D. A., Kramer, D. & Ceder, G. Non-equilibrium crystallization pathways of manganese oxides in aqueous solution. *Nat. Commun.* **10**, 1 (2019) doi: 10.1038/s41467-019-08494-6.
2. Baek, W., Das, S., Tan, S., Gavini, V. & Sun, W. Quasicrystal stability and nucleation kinetics from density functional theory. *Nat. Phys.* **21**, 980–987 (2025). doi: 10.1038/s41567-025-02925-6
3. Wang, F. & Landau, D. P. Efficient, multiple-range random walk algorithm to calculate the density of states. *Phys. Rev. Lett.* **86**, 2050 (2001) doi: 10.1103/PhysRevLett.86.2050.
4. Balluffi, R. W., Allen, S. M. & Carter, W. C. *Kinetics of Materials*. *Kinetics of Materials* (2005). doi: 10.1002/0471749311.